# Strength stimuli on the quadriceps femoris influences the physiological tremor of the index finger

Miguel J.A. Láinez[1], Luis-Millán González[2], José E. Gallach[2], Gerard Moras[3], Daniel Ranz[3], and José L. Toca-Herrera[4*]

[1] Departamento de Neurología, Hospital Clínico Universitario, Universidad de Valencia, Av. Blasco Ibáñez 17.46010 Valencia, Spain

[2] Departamento de Educación Física y Deporte, Universidad de Valencia, Gascó Oliag 3, 46010 Valencia, Spain

[3] Institut Nacional d'Educació Física de Catalunya, Av. de l'Estadi s/n, 08038 Barcelona, Spain

[4] Unidad de Biosuperficies, CIC biomaGUNE, Paseo Miramón 182, 20009 San Sebastián, Spain

**\*Corresponding author:**

José Luis Toca-Herrera

e-mail: jltocaherrera@cicbiomagune.es

Tel.: +34 943 005 313

Fax: +34 943 005 301

**ABSTRACT**

The influence of fatiguing stimuli applied to the quadriceps femoris on the tremor of the index finger on sixteen healthy subjects has been investigated, by measuring the acceleration and the electromyography activity of the extensor digitorum at two different speeds.

No significant changes in the basal condition could be measured at $0.52$ rad·s$^{-1}$. However, at $1.05$ rad·s$^{-1}$ the tremor amplitude ($12.23\%$, $p<0.05$) and the electromyography activity ($20.19\%$, $p<0.05$) increased in the time domain. In the frequential domain, the peak associated to the acceleration increased in $26.30\%$ ($p<0.05$), while the electromiography activity experienced an increase of $81.34\%$ ($p<0.05$). The largest change in frequency took place in the range 7-17 Hz. The discussion of the results leads to the conclusion that the supraespinal mechanism is responsible for the measured effect.

## <u>INTRODUCTION</u>

Tremor has been defined as a series of quick alternating involuntary rhythmic movements of contraction and relaxation which affect one or several parts of the body.[1] Historically, many authors have been interested in the mechanisms that underlie tremor.[2] Possibly, the extent of physiological tremor may be derived from the interaction of various neurogenic and mechanical factors.[3] Experimental evidence points out that the origin of physiological tremor might be due to the multifactorial action of central nervous system (CNS) structures,[4,5] reflex loops,[6] motor unit activity,[7-9] and of the mechanical characteristics of various joints.[10,11]

This *biological noise*[12] is considered an intrinsic property of the motor system normal performance, and is differentiated by various authors from other pathological forms such as essential, orthostatic and Parkinson's tremors, amongst others, in accordance with the frequency and amplitude at which they are produced.[13,14]

Studies on the frequency content of signals registered from the physiological tremor during the stable position of the upper limbs, show tremor frequencies with its highest values between 8 and 12 Hz.[15] These frequencies and their amplitude change depend on the position maintained by the assessed limb. Some authors suggest that strength training and moderate exercise influence postural tremor,[16,17] although certain controversy exists with regard to their effects.[18-20] Nonetheless, muscular fatigue, understood as a decline of muscular performance during exercise,[21] has an influence reducing the strength production capacity, increasing its variability (i.e. force fluctuations) and the postural tremor amplitude.[22-25]

A recent study has quantified the physiological tremor magnitude after distal body segment training.[26] This work explores the possibility that tremor between a limb and its contralateral side, are dependent. This relation would suggest that tremor increase may be due to changes in the supraspinal output signal. Nevertheless, it is known that unilateral training may influence the inactive side reflex mechanisms (e.g. cross extension reflex).[27] This so-called "cross education" phenomenon may be involved in the physiological tremor alterations produced in the contralateral homologous member. Based on this, our study objective is to establish the influence of a fatiguing stimulus applied to the femoral quadriceps on the index finger postural tremor. The choice of the lower limb as a training objective was to minimize the effects that may be produced on the peripheral and mechanical tremor factors associated with the upper limb in which the tremor has been assessed.

## METHODS

### *Subjects*

Sixteen healthy adult volunteers [ten males and six females: mean ± standard error of the mean (SEM); age 22.5±0.8 years; weight, 69.4±2.5; height 173.0±1.6 cm] took part in this research work. The volunteers usually practice moderate physical activity and none of them had any type of neurological, sensorial or physical disorder which could affect this study. These subjects were right-handed (kicking preference). None of them was taking medication. All were warned not to take stimulants (mainly caffeine derivatives) 24 h prior the measuring sessions. The research committee of the University of Valencia approved this study, and all subjects signed an agreement document before starting the test.

### *Experimental procedures*

Twenty-four hours after getting familiar with the different research protocols, the subjects underwent two sessions with a 48-hour period between both. In each session, the subjects performed a different isokinetic strength task, taking the measurements immediately before and after the task.

### *Mechanical records*

A 6.5 g accelerometer (K-Beam 8305A, Kistler, Amherst, USA) was used to measure postural tremor, for which 500 mV correspond to the acceleration of 1 $g$.

The accelerometer was placed at the 3rd phalange of the index finger (right hand), with the thumb in adduction while the remaining fingers were in flexion. The accelerometer was set in place with double-sided tape, and it was placed while the hand was in pronation, the elbow joint was fully extended and the flexion angle of the arm was 1.57 rad with regard to the shoulder. The subjects maintained the aforementioned arm position for 30 s while they were seated, without leaning their back against the chair back. Eight measurements were taken immediately prior to the isokinetic session and a further 8 measurements were taken right after this task.

The accelerometer signal, collected by means of an amplifier (gain x 10), was sampled at a rate of 1 kHz and was converted from analogical to digital (12-bit; DAQCard–700; National Instrument, Austin, USA).

A previous reproducibility test of the results was performed for this protocol by taking two separate measurements with a 48-hour period between them. Intraclass correlation coefficients for the test revealed good test–retest reliability (range, 0.83–0.89).

*Electric records*

During tremor measuring attempts, synchronized electromyography (EMG) activity of the extensor digitorum (ED) communis surface was registered. For this purpose, pre-gelled bipolar Ag/AgCl electrodes with a diameter of 10 mm were used (Blue Sensor M-00-S, Medicotest, Ølstykke, DNK). The electrodes were placed at inter-electrode distance 20 mm, and they coincided approximately with the center of ED muscle belly, roughly at this muscle's motor point. The reference electrode was placed on inactive olecranon tissue of the same arm being assessed (see Figure 1).

The EMG signals were sampled at a frequency of 1 kHz for each attempt. The EMG signal was amplified (gain x 5000). In the same way as the acceleration signal, the EMG signals were converted to digital from analogical and saved on a hard disk for future analysis.

*Isokinetic Protocol*

A CYBEX 6000 calibrated isokinetic dynamometer was used to perform the strength session (CYBEX Division LUMEX, Inc., Ronkonkoma, NY, USA). The subjects remained seated during the strength protocol with their arms relaxed, and their waist and trunk touching the chair back while they followed the manufacturer's instructions.[28] The training was performed on the right leg, within a flexion-extension range of the knee between 0 rad and 1.57 rad. The subject's leg was placed in line with the articulated machine arm so that the axis of the lever was aligned with the knee joint. The subjects applied strength on one roller at approximately the height of the fibula malleolus.

A standard warm-up session consisting of 6 repetitions of progressive intensity of knee flexion/extension at the programmed session speed was carried out prior to starting the protocol, The training, a unilateral isokinetic strength task of the knee flexors/extensors, consisted of two series of 30 s with a 1-minute rest in between. In the first session, the subjects did a maximum of 10 repetitions of knee extensors and another 10 knee flexors at a speed of 0.52 rad·s$^{-1}$. In the second session 48-hr later, a maximum of 20 repetitions of knee extensors were performed, as were a further 20 knee flexors at a contraction speed of 1.05 rad·s$^{-1}$. The parameters used in the training protocols (high intensity exercise in short time) have also been used in other studies, inducing neuromuscular alterations[29,30].

*Signal analysis*

Both the acceleration (tremor) and the surface electromyographic activity (EMG) signals were analyzed in the time and frequency domains. All signals were previously filtered digitally (high-passed and low-passed filter using an FFT-inverse FFT method) with a 1-50 Hz and 10-400 Hz, respectively. All analyses were carried out with Matlab 7.0 version software (MathWorks Release 14).

The first five seconds of the signal were not taken into account, and a total period of 25-s was analyzed. Tremor and EMG amplitude in the time domain were quantified by means of the root mean square (RMS), and processed every 100 ms.

The Matlab SPECTRUM function was used for the power spectral analysis of the EMG and tremor signals applying Welch's average. The analysis was performed using a window size of 512 data points.

The peak power (PP) was obtained for each spectrum quantifying the frequency associated with the peak power (PPF). In addition, the mean power peak of the whole spectrum (MPP) for four bandwidths was performed: 1-7, 7-17, 17-30 and 30-40 Hz.[26] A typical example of this analysis is shown in Figure 2.

*Statistical methods*

The first 8 attempts (pretest) average and also the last 8 attempts (post-test) average of the collected data from the EMG and the acceleration time domain, expressed as μV and $ms^{-2}$, were calculated. Likewise, the same mean was calculated for the frequency domain data. All values are plotted as a mean of $n = 16$ individuals.

Conventional statistical methods were used to calculate the mean and the standard error of the mean (SEM). Differences among means before and after training were found with a *t*-test for paired samples for those variables related to tremor and EMG activity (spectrum frequency, amplitude and the RMS). The probability of such differences being randomly caused were below 5% ($p < 0.05$) for all analyses.

**RESULTS**

Figure 3 illustrates typical tremor signals (acceleration) and EMG activity of the index finger and the ED respectively, both before and after training. When subjects underwent the isokinetic strength session with their right leg at a speed of 0.52 rad·s$^{-1}$, no changes were noted in the RMS of the postural tremor, and the same occurred with the EMG activity of the forearm under study.

Nevertheless, when the isokinetic strength protocol was carried out at a higher contraction speed (1.05 rad·s$^{-1}$), the acceleration produced in the index finger increased by 12.23% ($p<0.05$) in relation to the pretest. Likewise, considering specifically the amplitude of EMG activity of the ED, an increase of 20.19% ($p<0.05$) was obtained in the RMS after training. The results of the different trials and the mean of the pre and post-test are illustrated in Figure 4.

In the frequency domain analysis we saw that the frequency associated with the peak power frequency (PPF) underwent no significant changes under the two experimental conditions for both tremor and EMG. The PPF values ranged between 8.82-14.42 Hz and 68.13-196.39 Hz for EMG activity. In Table 1, the mean values of the PPF variable are shown.

No significant change was observed in the peak power variable (PP) when subjects were submitted to a fatiguing protocol at a speed of 0.52 rad·s$^{-1}$. On the contrary, at higher speeds the subjects presented percentage increases of 26.30% ($p<0.05$) in the PP variable. This was also noted in EMG activity, specifically, in the observed change was 81.34% ($p<0.05$).

The changes in the MPP variable for the different frequency bands may be seen in Figure 5. With regard to the percentage of the acceleration signal power, approximately the 66 % is seen to develop in the strip between 7-17 Hz for both speeds of 1.05 rad·s$^{-1}$ and 0.52 rad·s$^{-1}$, before and after training. In the EMG signal between 90 and 150 Hz, we may also observe ~70% of spectrum power for both speeds.

## DISCUSSION

Some theoretical models[31] base postural tremor control of the upper segments on the mechanical elements of the segment itself being assessed, the active elements which produce strength (i.e. the muscles themselves), and nerve factors (both spinal and supraspinal). Our work protocols have been directed to muscular fatigue on leg muscles to reduce the incidence of fatigue on mechanical and spinal factors, which may affect index finger tremor production. Historically speaking, potential factors which contribute to muscular fatigue include central and peripheral factors. The former could cause fatigue through the distortion of neuromuscular transmission between the central nervous system and the muscle membrane, whereas the latter could alter the muscle membrane.[32]

The fact that a fatiguing exercise affects postural tremor has been proved in numerous research works.[22,23] In this sense, we have observed significant changes in our study in the amplitude of the index finger tremor and the EMG activity after a training session involving the right leg when subjects performed a protocol at a speed of 1.05 rad·s$^{-1}$. In relation to the pre-test, the percentages of increase were around 12% in tremor and 20% in EMG, which were slightly lower than those found by Morrison and collaborators.[26] These authors established that unilateral (fatigue) training not only affects the tremor in the exercised segment but also the contralateral segment tremor. The fact that in this work wrist extensors and flexors are innerved at the same medullar level ($C_5$-$T_1$) as the contralateral muscles responsible for the control of the index finger tremor, may have influenced the specific reflex responses which magnify such changes.

Apparently, the alterations which took place at the speed of 1.05 rad·s$^{-1}$ in the amplitude of tremor are modulated by a central type fatigue, as confirmed by the tremor signal analysis. The changes in the spectrum amplitude at low frequency range (7-17 Hz) are representative of the supraspinal oscillations produced in the upper centers of the central nervous system, which include the motor cortex and the thalamus.[33-35]

We have also observed lower changes in the tremor amplitude associated at high frequencies (17-30 Hz) probably due to muscular fatigue caused by the 8 repetitions that took place in the pre- and post-test. Figure 4 shows that no significant changes (that means no fatigue) were measured in the pre-test try, meaning that the experimental protocol was correct.

Similar features in the frequency analysis appeared in the EMG activity registered in the ED, in which the PPF variable remained at the same frequency range whereas the PP-associated amplitude increased significantly. Other studies that are not directly comparable with our work reported the decrease of such frequencies after extenuating strength protocols[34]. In our study the number of repetitions (between 10 and 20) were not enough to change the PPF profile.

The two contraction speeds used in our study show different behaviors. For a speed of 0.52 rad·s$^{-1}$ no significant changes in tremor are observed because the greatest affectation is noted at a peripheral level of the central nervous system (i.e. fatigue produced at a local level), where no tremor control mechanisms of a remote limb are altered.[37,38] This is reported in some investigations, which have compared low charge at high speed with high charge at low speed protocols. Large local accumulations of muscle lactic acid have been found for high charge at low speed protocols.[39]

In this way, only the index finger tremor is affected by the exercise of a remote limb when the charge magnitude is large enough to influence central nervous system supraspinal processes.

The present work asserts that postural index finger tremor is altered after training with a remote and distal limb. This may be due to an intervention of the supraspinal mechanisms in the process.

Our results also point out that the organism would show larger fatigue when muscular work is performed at high speeds in short time.

**Figure captions**

**Figure 1.** Image of how the different sensors used in the experiment were placed.

**Figure 2.** Spectrum amplitude of both the acceleration signal (tremor) and EMG of a typical subject before and after a completing an exercise protocol. The broken line represents the amplitude in the pretest while the continuous line represents the post-test.

**Figure 3.** Example of tremor signals (acceleration) of the index finger and the EMG activity of the ED (forearm), measured before and after performing the isokinetic strength training with one leg at a contraction speed of 1.05 rad·s$^{-1}$. The duration of both signals was 30 s.

**Figure 4.** Changes in the mean index finger tremor and in the EMG activity of the ED before and after performing the fatiguing exercise protocol at 1.05 rad·s$^{-1}$. Left Figures show the mean value of the different trials, squares and circles represent respectively the pre-test and post-test values. Right Figures show the mean value of the eight pre-test and post-test trials. Note that each value is based on the mean values for all subjects (n=16). Error bars represent the SEM. Significant differences between these values are marked with an * (p<0.05).

**Figure 5.** Mean ranges of tremor before and after isokinetic strength training for the different frequency strips. Each value is based on the mean values for all subjects (*n=16*). Bars represent the SEM. Significant differences between values are marked with an * (p<0.05).

**Figure 1**

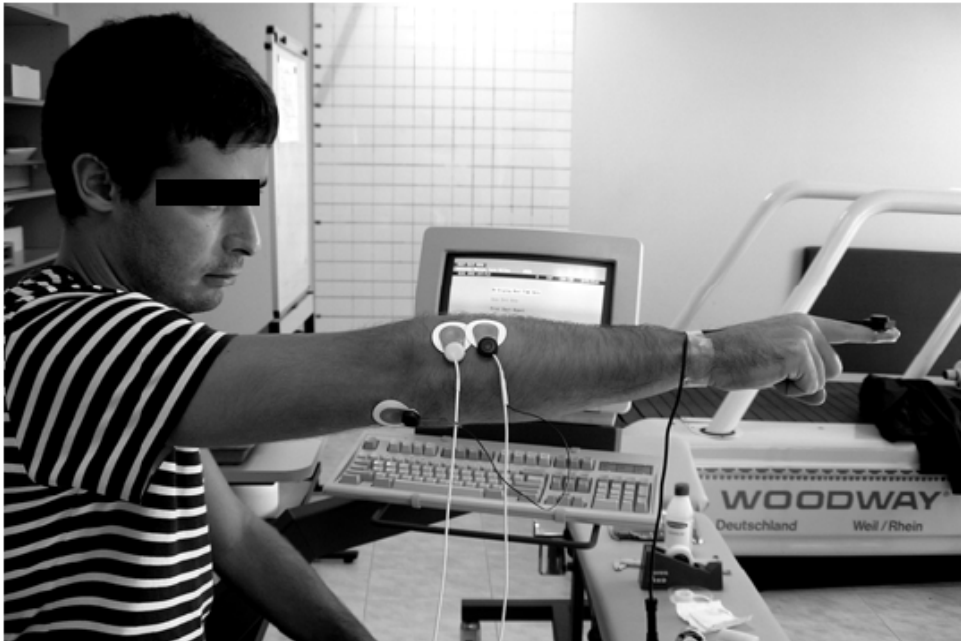

**Figure 2**

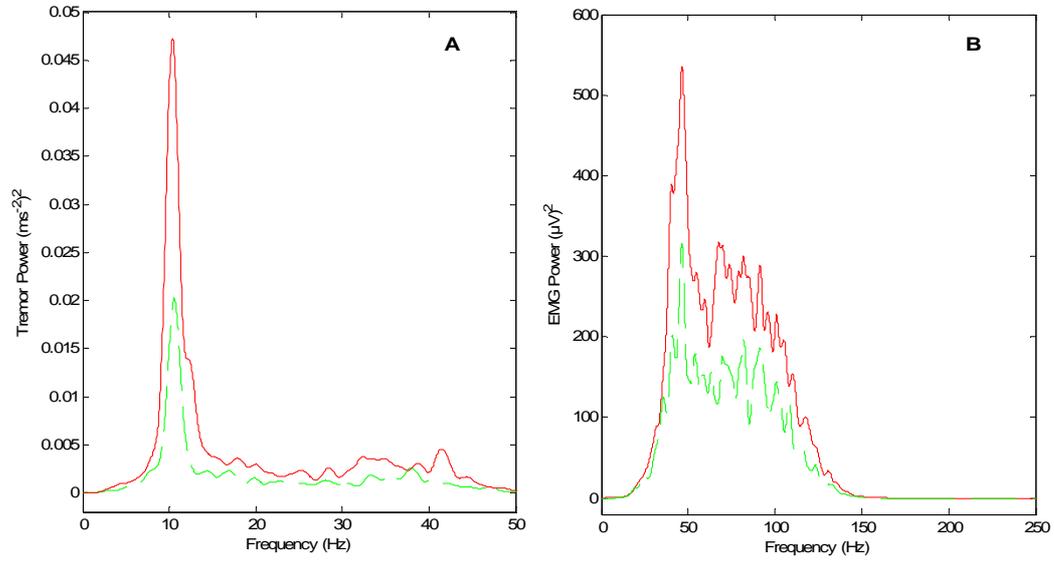

**Figure 3**

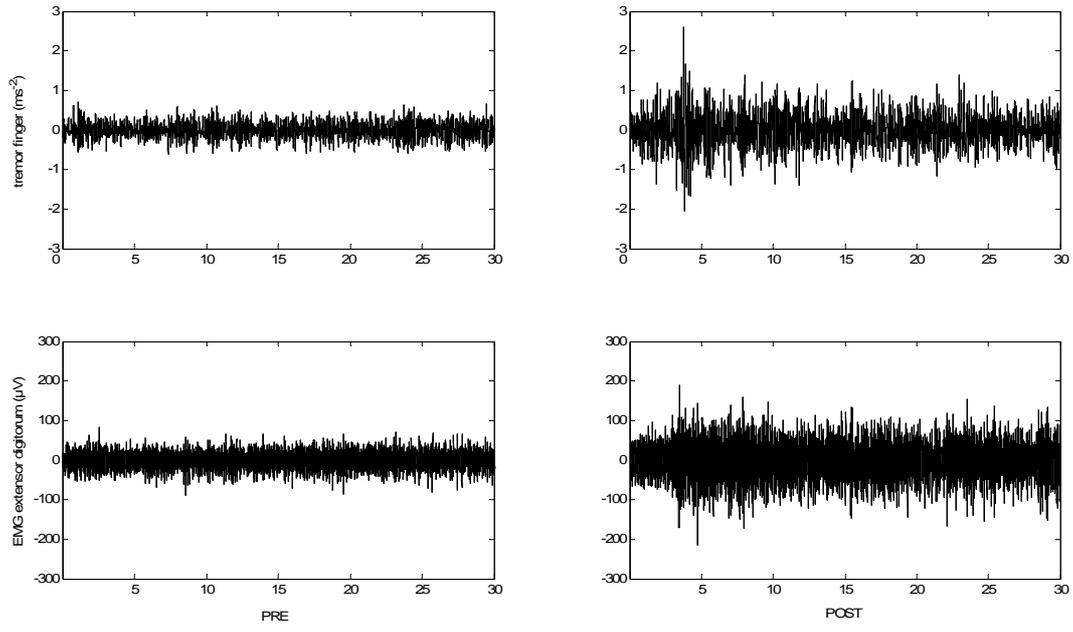

**Figure 4**

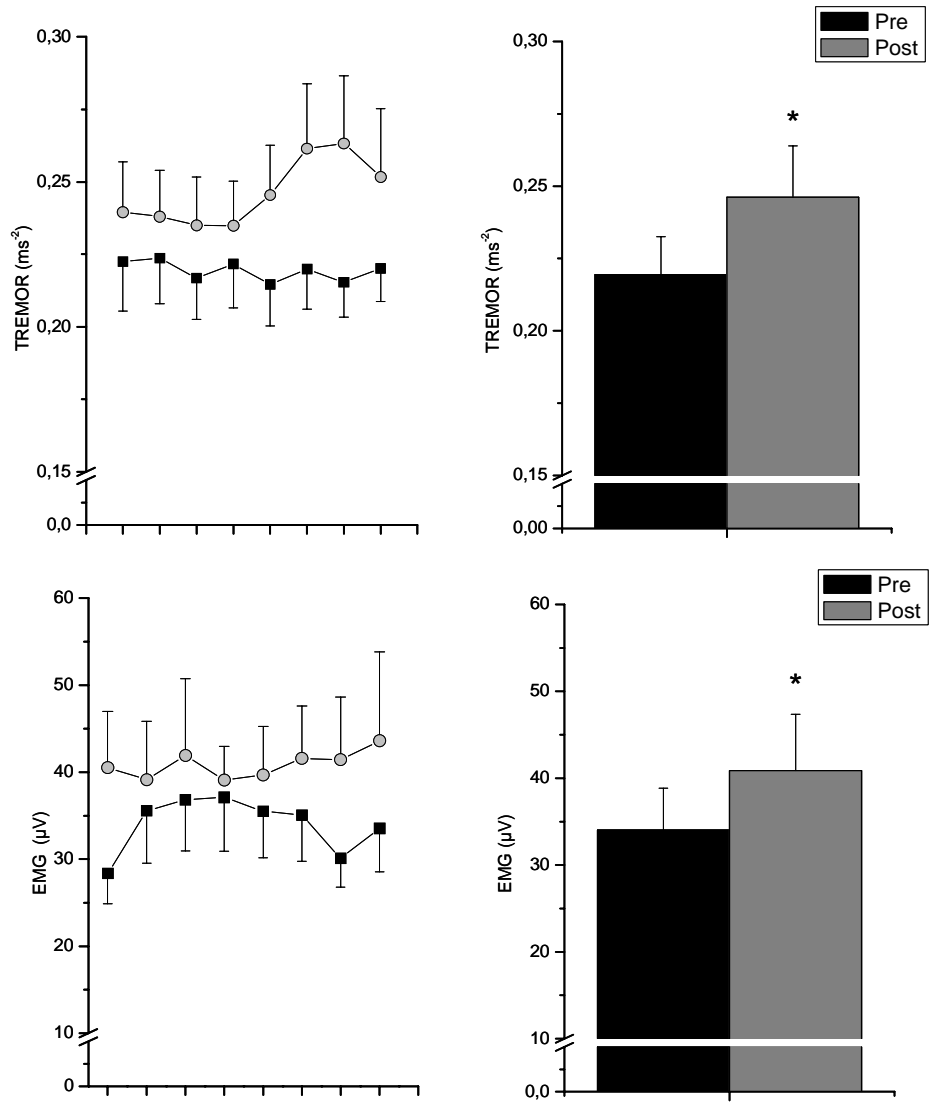

**Figure 5**

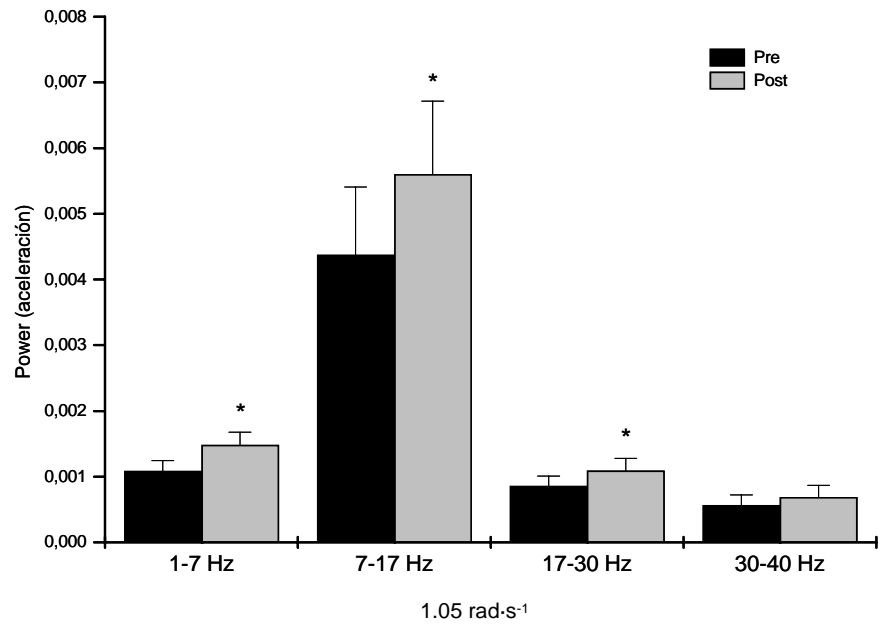

**Table 1.** Mean values of the PPF variable of both electromyographic records and tremor. Data are expressed in Hz (mean value ± SEM).

| PPF | $0.52 \ rad \cdot s^{-1}$ | | $1.05 \ rad \cdot s^{-1}$ | |
|---|---|---|---|---|
| | **Pre** | **Post** | **Pre** | **Post** |
| EMG | 89.43±3.36 | 108.93±9.35 | 91.41±4.01 | 106.61±8.15 |
| TREMOR | 10.41±0.32 | 11.64±1.44 | 9.97±0.24 | 10.10±0.26 |